# Characterization of Multi-Core Fiber Group Delay with Correlation OTDR and Modulation Phase Shift Methods

Florian Azendorf[1,3], Annika Dochhan[1], Krzysztof Wilczyński[2], Łukasz Szostkiewicz[2], Patryk Urban[2], Bernhard Schmauss[3], Francisco Javier Vilchez[4], Laia Nadal[4], Michela Svaluto Moreolo[4], Josep M. Fabrega[4], Michael Eiselt[1]

[1]*ADVA Optical Networking SE, Maerzenquelle 1-3, 98617 Meiningen, Germany*
[2]*InPhoTec Sp. z o.o., Meksykańska 6 lok. 102, 03-948 Warsaw, Poland*
[3]*LHFT, Friedrich-Alexander Universität Erlangen-Nürnberg, 91058 Erlangen, Germany*
[4]*Centre Tecnològic de Telecomunicacions de Catalunya (CTTC/CERCA), Castelldefels (Barcelona), Spain*
*Fazendorf@advaoptical.com*

**Abstract:** Using a Correlation-OTDR and a modulation phase shift method we characterized four multi-core fibers. The results show that the differential delay depends on the position of the core in the fiber and varies with temperature.

## 1. Introduction

The propagation delay of a fiber link becomes a critical parameter for future 5G networks. Not so much the absolute delay value is of a concern, but rather the differential delay between two optical paths. The application of analog radio over fiber and the use of phase array antennas permit a maximum phase error of 30 degrees at 26 GHz. This corresponds to a maximum skew of 3.2 ps. Another area where the differential delay is important are synchronization applications. In current synchronization applications, like IEEE 1588v2 (PTP), an asymmetry of a few nanoseconds between the master and slave clocks is sufficient. Future synchronization applications might require an asymmetry reduced by an order of magnitude. One of the main causes of propagation delay changes is temperature. For a standard single mode fiber, the temperature delay coefficient (TDC) is 7.49 ppm/degC [1]. To reduce differences between two optical paths, instead of using two different fibers within one cable or multiple wavelength in one fiber multi-core fibers might be beneficial. To get an insight into the temperature dependent skew of multi-core fibers, the cores must be characterized simultaneously. Temperature changes can occur even within a few minutes, which impact the measured fiber delay and could lead to an incorrect value for the skew between the cores, if not measured simultaneously. In this paper, we report on the characterization of four different multi-core fiber samples, two 7-core fibers and two 19-core fibers. All fibers mentioned in this paper are characterized with two measurement methods. The first method uses the modulation phase shift (MPS) technique, which is well-known to measure the chromatic dispersion (CD) as well as the DGD of an optical fiber and is implemented in commercial measurement equipment [2]. This technique requires access to both ends of the fiber, and each fiber core is measured consecutively. For the applications mentioned above, a second method based on correlation optical time domain reflectometry (C-OTDR) was developed to measure the propagation delay in a fiber from one fiber end with an accuracy in the order of few picoseconds [3]. With this method we can characterize four cores simultaneously.

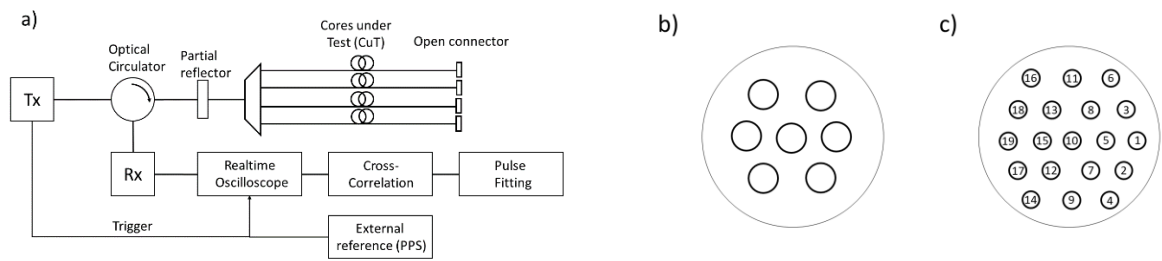

Fig. 1. a) Schematic of the C-OTDR to measure four cores simultaneously, b) 7-core fiber, two different vendors, 1 km/10 km c.) 19-core fiber, two different vendors, 5 km/25 km

## 2. Experimental setup

The measurement setup of the C-OTDR is shown in Fig. 1a. A continuous wave laser signal with a wavelength of 1550 nm was modulated, using a Mach-Zehnder modulator. We used complementary 2048-bit Golay sequences [4] followed by a period of constant power with a duration depending on the fiber length. The signal was sent through an optical circulator and was reflected at a connector with an air gap, providing a reference reflection at the input to the fiber. To measure several cores simultaneously, the signal was split using a 1x4 coupler. As the fiber output was unterminated, we obtained a reflection of approximately 4% from the end surfaces of each core. The combination of the reflections from all cores under test were received with a PIN/TIA combination and sampled on a real-time oscilloscope with 50 GS/s. 4000 signals were recorded and averaged for each trace. The real-time oscilloscope was synchronized by a trigger from the burst source. A precise external clock, based on an oven-controlled crystal oscillator (OCXO) with a stability of 5 ppb and linked to five GNSS satellites, was used for a long-term stable timing accuracy. The averaged signal is shown in Fig. 2a. Due to the data rate of 10 Gbit/s the pulse width of one burst is 204.8 ns, and the four reflected signals are linearly superimposed. Due to the similar propagation times in all cores, the reflections are not distinguishable within this averaged signal. The averaged signal was then cross-correlated with the transmitted sequences, and the correlation results of the complementary sequences were added. After the cross correlation, the correlation peaks from the different cores are clearly distinguishable, separated by up to 4 ns, as shown in Fig. 2b. Due to the sampling rate of 50 GS/s the time-resolution was 20 ps. To achieve an accuracy of better than one sample period, a Gaussian function was fitted to the correlation values around the highest reflection peak, as shown in Fig. 2c.

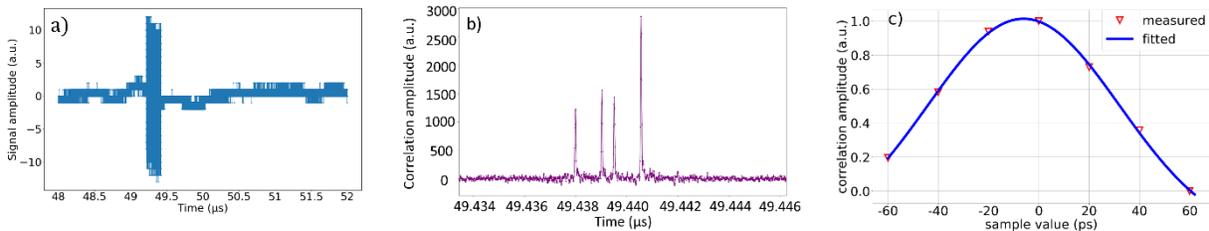

Fig. 2. a) Averaged received signal showing reflected Golay sequences, b) sum of cross-correlation between received signal and Golay sequence, c) fit of one correlation peak to a Gaussian function.

## 3. Results

### 3.1. PMD measured with MPS-method

Using the MPS method, the DGD for each core of the fibers under test was measured over a spectral bandwidth of 110 nm with a resolution of 0.05 nm. The PMD was calculated as the average value of the DGD. The results show high PMD values (up to 22 ps) for either only the outer cores or all cores of the fiber, except the central core, which shows in all cases PMD values below < 0.43 ps. We attribute the high PMD values to stress induced birefringence. In addition to the center cores, for the 19-core fiber, low PMD values are measured for the first core ring with neighbors around them, whereas the corner cores, as an example core 16 and core 6 in Fig. 1c, exhibit PMD values up to approximately 6 ps. This might be attributed to stress induced by the neighbor cores. The maximum and minimum core PMD values as well as the average over all cores are shown for all fibers in Table 1.

Table 1. Length, PMD maximum, minimum, average and delay difference maximum and minimum of different multi-core fibers.

|  | a) 7-core fiber | b) 7-core fiber | c) 19-core fiber | d) 19-core fiber |
|---|---|---|---|---|
| length (km) | 10 | 1 | 5 | 25 |
| PMD min/max (ps): | 0.68/22.1 | 0.26/0.77 | 0.16/5.98 | 0.25/1.19 |
| PMD average (ps); | 7.76 | 0.51 | 1.81 | 0.58 |
| delay difference min / max (ns) | -1.77/3.81 | -1.2/0.33 | -3.31/1.88 | -31/28 |

### 3.2. Propagation delay measurements

Using the C-OTDR, the propagation delay of all cores under test was measured. In each measurement, four cores were characterized simultaneously, one of those being the center core of the fiber as reference for the differential delay, or skew, calculation. The maximum and minimum delay differences for all fibers are shown in Table. 1. The measured skew between the fiber cores was as large as 25 ns. It was observed that the skew between the cores depends on the position of the core in the fiber and the fiber length. One side of the fiber cross area experienced a higher delay than the opposite side, which might be explained by spooling of the fiber. During the measurements of a 19-core 25 km fiber, delay differences from 10 ns to 31 ns were obtained, which might be caused by the stress in the outer cores induced by the spooling of the fiber. To evaluate the accuracy of the C-OTDR measurement method, we also measured the delay of the cores with the MPS method. The difference between both methods yielded a constant offset of up to approximately 300 ps in case of the 5-km 19-core fiber. This difference might result from the fact that the time base of the MPS equipment was not synchronized to the high-precision clock used for the C-OTDR.

### 3.3. Temperature dependent propagation delay measurements

The test fibers were placed into a temperature-controlled cabinet, and the propagation delay was measured over a temperature range of 10°C to 50°C in steps of 10 K. A temperature settlement time of 1.5h was allowed for every temperature step. Fig. 3. shows the evolution of the fiber group delay over temperature for a 10-km 7-core fiber and a 5-km 19-core fiber. All fibers showed a relatively linear behavior over temperature. This behavior is comparable to a bare single mode fiber with a constant temperature delay coefficient (TDC) of 7.1 – 7.49 ppm/K.

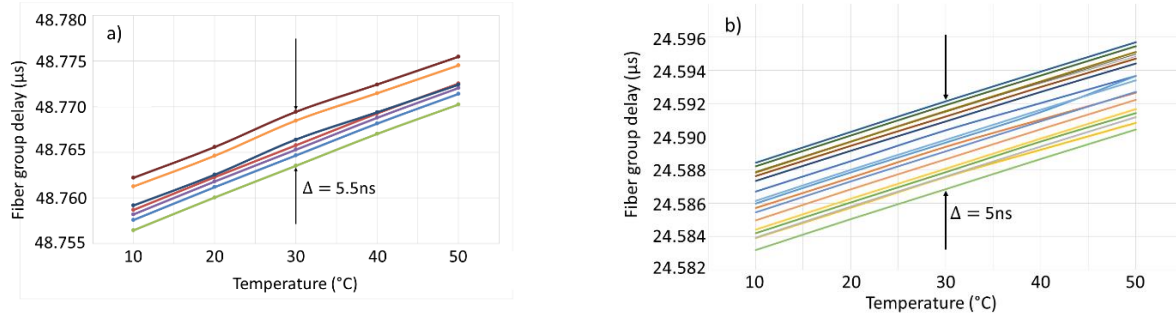

Fig. 3. Group delay evolution over temperature of a) 10-km 7-core fiber, b) 5-km 19-core fiber.

### 3.4. Skew over temperature with respect to the center core

Important information for synchronization applications is the temperature dependent skew between different fiber cores. With the results of the temperature dependent propagation delay measurement we calculated the skew over temperature with respect to the central core. While over a temperature range of 40 K the skew changes between 1 ps and 35 ps, no consistent trend could be observed. We attribute these variations to the fiber PMD, which would lead to a random differential group delay of several times the PMD value.

### 4. Conclusion

Using a Correlation OTDR with an accuracy of approximately 3 ps and, alternatively, the modulation phase shift (MPS) method, we characterized the delay and PMD of four multi-core fibers with different lengths. While one fiber showed high PMD values of all cores, another fiber showed high PMD values only at the corner cores. All the fibers showed cores with positive or negative delay differences with respect to the center core. We assume that the reason for the delay differences is the spooling of the fiber. These values give an insight how the cores are distributed in the cross section. While static delay difference can be compensated upon deployment, the delay variations (skew) over temperature require a precise monitoring and can only be compensated, if they are not based on PMD. For novel synchronization applications, multi-core fibers might be used, if the delay differences are monitored with a precise accuracy.

### 5. Acknowledgment

The authors thank Dr. Ming-Jun Li for providing two of the multi-core fibers under test. This work has received funding from the European Union´s Horizon 2020 research and innovation programme under grant agreement No 762055 (BlueSpace Project).